\documentclass[11pt]{article}

\usepackage{color}
\usepackage{graphicx,times,amssymb}
\usepackage{natbib}

\def\aj{AJ}%
\def\araa{ARA\&A}%
\def\apj{ApJ}%
\def\apjl{ApJ}%
\def\aap{A\&A}%
\def\azh{AZh}%
\def\jcap{J. Cosmology Astropart. Phys.}%
\def\mnras{MNRAS}%
\def\nat{Nature}%
\def\bain{Bull.~Astron.~Inst.~Netherlands}%
\input epsf.sty

\begin{document}

\title{Dark Matter}

\author{Jaan Einasto\\
Tartu Observatory, ICRANet}
\maketitle
\begin{abstract}
  I give a review of the development of the concept of dark matter.
  The dark matter story passed through several stages from a minor
  observational puzzle to a major challenge for theory of elementary
  particles.  Modern data suggest that dark matter is the dominant
  matter component in the Universe, and that it consists of some
  unknown non-baryonic particles.  Dark matter is the dominant matter
  component in the Universe, thus properties of dark matter particles
  determine the structure of the cosmic web.
\end{abstract}

\section{Dark matter story}

The masses of astronomical bodies are usually determined directly,
using motions of other bodies around or within the body under
study. In some cases total mass estimates found by different methods,
differ by a large fraction. It is customary to call the hypothetical
matter, responsible for such mass discrepancy, dark matter (DM).

The timeline of the study of dark matter is shown in Table
\ref{Tab1}. Actually there are two dark matter problems --- the local
dark matter close to the plane of our Galaxy, and the global dark
matter surrounding galaxies and clusters of galaxies.  However, this
difference was understood only later, thus we show in the Table the
whole story.

{\small
\begin{table}[ht]
\caption{Dark Matter Timeline.}
\begin{tabular}{ll}
\noalign{\smallskip}  
\hline\hline 
\noalign{\smallskip}  
Year  & Description \\ 
\noalign{\smallskip}
\hline
\noalign{\smallskip}
1915             &First estimates of local DM: \citet{Opik:1915},
\citet{Kapteyn:1922}, \citet{Jeans:1922fk} \\
1932               &Galactic model and Local DM:  \citet{Oort:1932}\\
1933               &DM in Coma cluster: \citet{Zwicky:1933}\\
1952             &Galactic models and local DM: 
\citet{Kuzmin:1952,Kuzmin:1952a,  Kuzmin:1955}\\
1957&Large $M/L$ on the periphery of M31:
\citet{van-de-Hulst:1957um}, \citet{Roberts:1966dt} \\
1959              &Mass of the Local Group: \citet{Kahn:1959}\\
1961              &Cluster Stability Conference: 
\citet{Neyman:1961zr}\\
1965 &Discovery of CMB: \citet{Penzias:1965} \\
1972              &Cluster X-ray data on mass of hot gas:  \citet{Forman:1972zr,Gursky:1972fr}\\
1972              &Local and global DM different, global DM  
non-stellar: \citet{Einasto:1972t,Einasto:1974a}\\
1974              &Parameters of DM coronas/halos:
\citet{Einasto:1974}, \citet{Ostriker:1974}\\
1975              &DM contradicts  classical cosmological
paradigm: \citet{Materne:1976}\\
1977              &$M/L$ of galactic bulges low: \citet{Faber:1977}\\
1978              &Extended flat rotation curves: \citet{Bosma:1978},
\citep{Rubin:1978}\\
1984              &Cold DM accepted:
\citet{Blumenthal:1984}\\
1989             &Absence of large amounts of local DM accepted:
\citet{Gilmore:1989}\\
2012             &CDM particle annihilation detected?:
\citet{Weniger:2012}, \citet{Tempel:2012fu}\\ 
\noalign{\smallskip}
\hline
\end{tabular}
\label{Tab1}                        
\end{table}
}

\section{Local dark matter}

The first indication for the possible presence of dark matter came
from the dynamical study of our Galaxy. \citet{Opik:1915} was probably
first to estimate the dynamical density of matter in the Galaxy in the
vicinity of Sun. \"Opik analyzed vertical motions of stars near the
plane of the Galaxy and calculated the dynamical density.  He also
estimated the density due to all stars near the Galactic plane using
the luminosity function of stars.  Dutch astronomer Jacobus
\citet{Kapteyn:1922} made a similar analysis. \"Opik and Kapteyn found
that the spatial density of known stars is sufficient to explain the
vertical motions.  British astronomer James \citet{Jeans:1922fk}
reanalysed vertical motions of stars near the plane of the Galaxy, and
found that some dark matter probably exists near the Sun.

A new model of the Galaxy was calculated by Jan Hendrik
\citet{Oort:1932}, who also determined the dynamical density of matter
near the Sun. Oort accepted as most probable value 0.092 Solar masses
per cubic parsec. He found the density due to visible stars --- 0.038
Solar masses per cubic parsec. This difference is often considered as
an indication for the presence of dark matter. Oort estimated
the total expected mass of faint stars, which is very near to the value
found from vertical motions of stars.

\citet{Kuzmin:1952,Kuzmin:1955} and his students in Tartu Observatory
showed that the amount of DM in the Galactic disk is small; in
contrast \citet{Hill:1960}, \citet{Oort:1960}, \citet{Bahcall:1984c}
and some other astronomers found evidence that up to a half of matter
in the Solar vicinity may be dark. More accurate recent data showed
that the amount of local dark matter is small \citep{Gilmore:1989}.
If there is some local dark matter, it must be dissipative to release
extra kinetic energy during the contraction of matter to a flat disk.
This population probably consists of very faint stars or Jupiter-like
objects.

\section{Global dark matter}

\citet{Zwicky:1933,Zwicky:1937uz} measured redshifts of galaxies in
the Coma cluster and found that the velocities of individual galaxies
with respect to the cluster mean velocity are much larger than those
expected from the estimated total mass of the cluster, calculated from
masses of individual galaxies. The only way to hold the cluster
from rapid expansion is to assume that the cluster contains huge
quantities of some invisible dark matter. According to his estimate
the amount of dark matter in this cluster exceeds the total mass of
cluster galaxies at least tenfold, probably even more. At this time 
astronomers were interested in the structure and evolution of stars,
and Zwicky's work seemed to be remote and uninteresting.
 
Slowly more dynamical data on clusters of galaxies were
collected, and the discrepancy between the cluster galaxy measured
velocities and expected velocities for a stable cluster could not be
ignored.  In 1961 during the International Astronomical Union (IAU)
General Assembly a special meeting to discuss the stability of
clusters of galaxies was organised by \citet{Neyman:1961zr}.  However,
opinions of astronomers were different and no definite conclusions
were achieved.

\citet{Kahn:1959} paid attention to the fact that most galaxies have
positive redshifts as a result of the expansion of the Universe; only
the Andromeda galaxy M31 has a negative redshift of about 120 km/s,
directed toward our Galaxy. This fact can be explained if both
galaxies form a physical system. From the approaching velocity, the
mutual distance, and the time since passing the perigalacticon (taken
equal to the present age of the Universe), the authors calculated the
total mass of the double system, about 5 times the sum of the
conventional masses of the Galaxy and M31. The authors suggested that
the extra mass is probably in the form of hot gas of temperature about
$5 \times 10^5$ K.

A similar problem exists in double elliptical galaxies. The mean
mass-to-luminosity ratio of double elliptical galaxies is $M/L
\approx$66 \citep{Page:1952,Page:1959}, much higher than estimated
masses of individual elliptical galaxies. A certain discrepancy was
detected also between masses of individual galaxies and masses of
groups of galaxies \citep{Holmberg:1937uq,Page:1960}.

\citet{Babcock:1939,Oort:1940, Roberts:1966dt} and \citet{Rubin:1970}
discovered that rotation curves of spiral galaxies are flat on the
periphery of galaxies. This is contrary to expectations since the
luminosity of galaxies falls rapidly on the periphery and a Keplerian
decrease of the rotation curve is expected. If rotation velocities are
identified with circular velocities, these observations suggested very
high values of mass-to-luminosity ratios $(M/L)$ on the periphery of
galaxies.

Detailed models of galaxies using available data on all basic stellar
populations (core, disk, bulge, halo, flat population of young stars
and interstellar gas) suggested that it is impossible to reproduce the
rotation curves of galaxies using independent data on $M/L$-ratios of
galactic populations \citep{Einasto:1965,Einasto:1969b,Einasto:1972}.
In order to bring rotation data into agreement with data on known
populations it is needed to assume the presence of a new population
with very large $M/L$-value, mass and radius
\citep{Einasto:1972t,Einasto:1974a}.   To find main parameters of the
new population --- corona --- the motion of satellite galaxies was
studied by \citet{Einasto:1974}.  This analysis showed that the mass
and effective (harmonic) radius of the corona exceeds the mass and
radius of known populations almost tenfold.  This analysis was
confirmed by \citet{Ostriker:1974}; both independent data suggested
that the previously unknown population dominates the mass budget of
the Universe: the mean density of matter is about 0.2 of the critical
cosmological density. Ostriker et al. used the term `halo' to denote
the massive population.

The dark matter problem was discussed in a regional conference on
January 1975 in Tallinn.  The main subject of the discussion was the
nature of the dark matter.  Two basic models were suggested for dark
matter: faint stars or hot gas.  It was found that both models have
serious difficulties.

The problem was discussed again at the Third European Astronomical
Meeting in Tbilisi on June 1975. This Meeting was the first well
documented international discussion between supporters and opponents
of the dark matter concept.  \citet{Materne:1976} concluded that
systems of galaxies are stable with conventional masses. Their most
serious argument was: Big Bang nucleosynthesis suggests a low-density
Universe with the density parameter $\Omega \approx$0.05; the
smoothness of the Hubble flow also favours a low-density Universe.

Soon new observational data arrived which supported the presence of
massive halos/coronae of galaxies.  Both optical
\citep{Rubin:1978,Rubin:1980}, and radio \citep{Bosma:1978} data
confirmed flat rotation curves of galaxies at large galactocentric
distances.  \citet{Faber:1976,Faber:1977} found that velocity
dispersions and mass-to-light ratios for elliptical galaxies and
bulges of galaxies  are considerably lower that expected
before.  This observation demonstrates clearly that conventional
stellar populations cannot be identified with dark halos/coronae. 

New independent observations confirmed large masses of clusters of
galaxies.  Already early X-ray observations of clusters of galaxies
showed that the mass of the hot X-ray emitting gas is not sufficient
to hold clusters together.  These data also allowed to estimate total
masses of clusters, and confirmed earlier measurements on the basis of
velocity dispersions of galaxies in clusters
\citep{Forman:1972zr,Gursky:1972fr,Kellogg:1973ly}.

Additional estimate of masses of clusters of galaxies came from
gravitational lensing, also supporting large masses of clusters of
galaxies. 

\section{The nature of dark matter}

By the end of the 1970's most objections against the dark matter hypothesis
had been rejected.  However, there remained three problems:

\begin{itemize}
  
\item{} It was not clear how to explain the Big Bang nucleosynthesis
  constraint on the low density of matter, and the smoothness of the Hubble
  flow --- the main argument in favour of the classical cosmological paradigm.
  
\item{} If the massive halo (corona) is neither stellar nor gaseous, of what stuff
  is it made of?
  
\item{} And a more general question: What is the role of dark matter
  in the evolution of the Universe?

\end{itemize}

Answers to these problems came from completely new areas of research
--- observations of the  cosmic microwave background (CMB) radiation,
and large-scale distribution of galaxies.

According to the current understanding, the Universe began with a Big
Bang and was initially very hot. It expanded rapidly and cooled, and
at a certain epoch was cool enough for atoms to recombine.  The effective
temperature of this radiation drops as the Universe expands.  The
cosmic microwave background radiation, a remnant of the initial
hot Universe, was detected by the American radio astronomers \citet{Penzias:1965}.

As emphasised by \citet{Peebles:1970} and \citet{Zeldovich:1970}
 structures in the Universe were created by the growth of
small inhomogeneities of the density. During the initial hot phase of
the evolution of the Universe the matter and radiation were coupled
and density inhomogeneities could not grow.  As the Universe expanded,
the gas cooled, and at recombination the gas became neutral.  From
this time on, density fluctuations in the gas had a chance to grow by
gravitational instability. The density fluctuations are of the same
order as temperature fluctuations.  Thus astronomers started to search
for temperature fluctuations of the CMB radiation. None were found,
only lower upper limits for the amplitude of CMB fluctuations were
obtained. On the other hand, theoretical calculations show that at the epoch of
recombination the density (and temperature) fluctuations must have an
amplitude of the order of $10^{-3}$.  Otherwise structure cannot
form, since the gravitational instability works very slowly in an
expanding Universe. 

\begin{figure*}[ht]
\centering
\resizebox{0.6\textwidth}{!}{\includegraphics{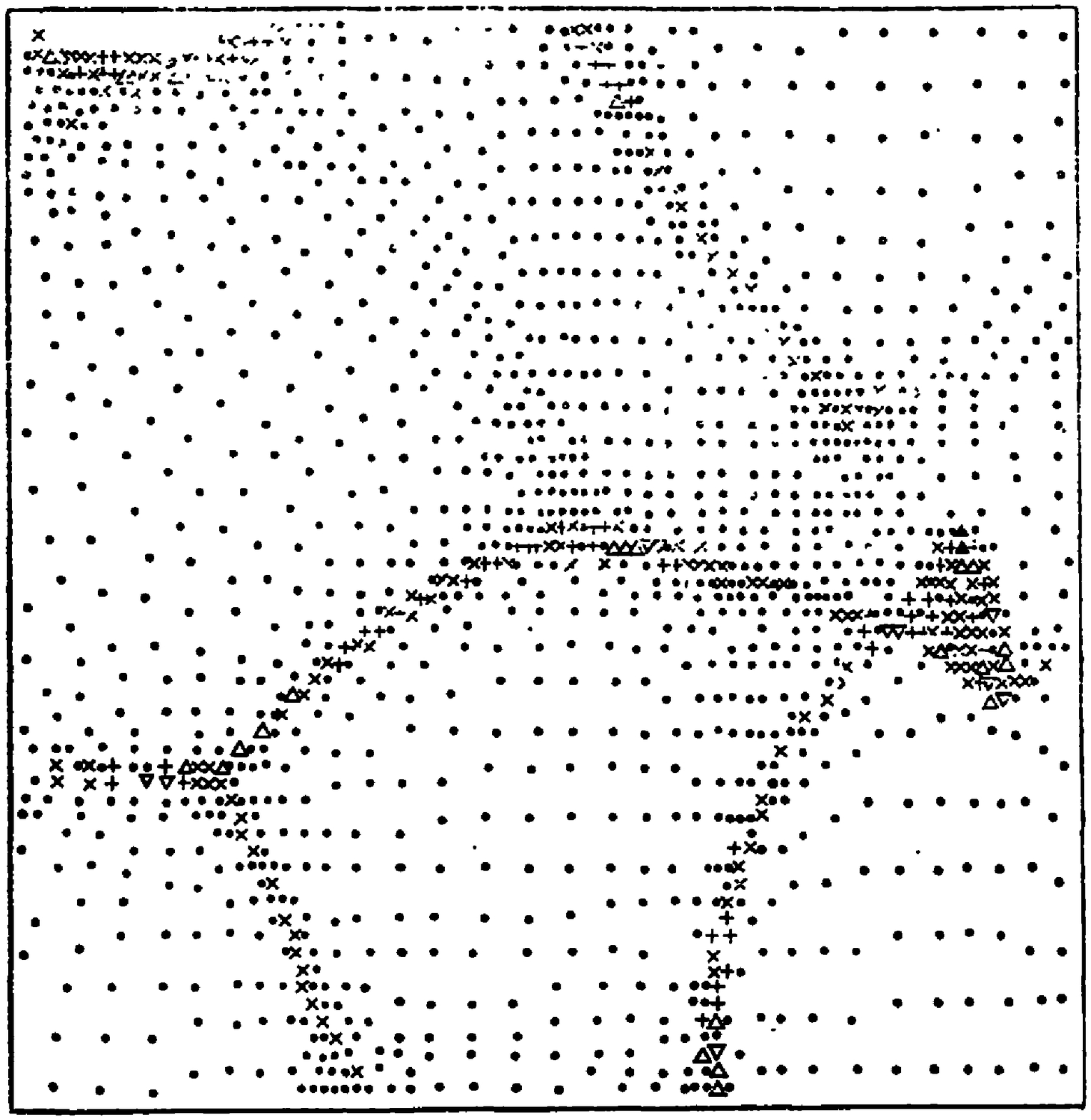}}
\caption{Distribution of particles in simulations
  \citep{Doroshkevich:1978vn}. }
\label{fig:model}
\end{figure*}

This controversy can be solved if non-baryonic elementary particles,
such as massive neutrinos, form dark matter particles.  There were
several reasons to search for non-baryonic particles as a dark matter
candidate.  First of all, no baryonic matter candidate fit the
observational data.  Second, the total amount of matter is of the
order of $0.2-0.3$ in units of the critical cosmological density,
while the nucleosynthesis constraints suggest that the amount of
baryonic matter cannot be higher than about 0.04 of the critical
density. If dark matter is non-baryonic, then this helps to explain the
paradox of small temperature fluctuations of the cosmic microwave
background radiation.  Density perturbations of non-baryonic dark
matter already start growing during the radiation-dominated era,
whereas the growth of baryonic matter is damped by radiation.

\begin{figure*}[ht]
\centering
\resizebox{0.6\textwidth}{!}{\includegraphics{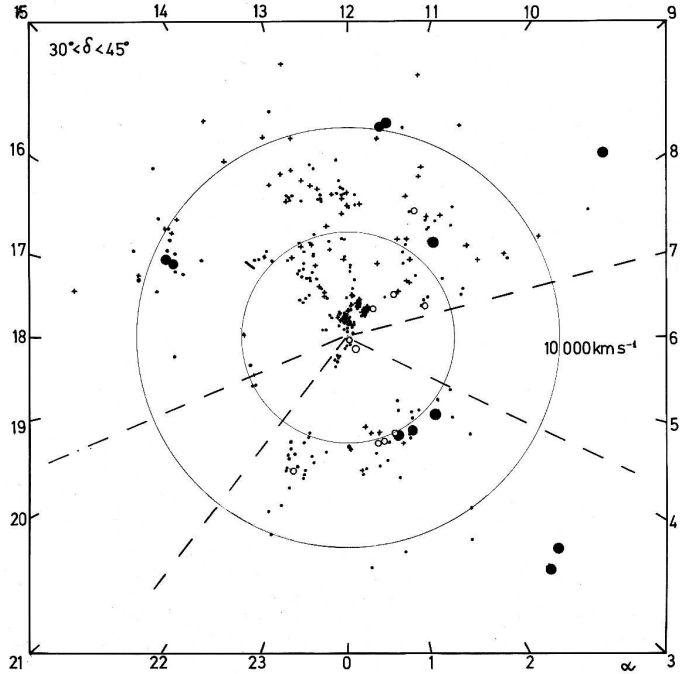}}
\caption{Wedge diagram for the $30^\circ-45^\circ$ declination
  zone. Clusters at RA about 2 h belong to the main chain of clusters
  and galaxies of the Perseus--Pisces supercluster; galaxies and
  clusters near the center at RA about 12 h are part of the Local
  supercluster, and galaxies and clusters at redshift about 7000 km/s
  and RA between 10 h and 13 h belong to the Coma supercluster. Note the
  complete absence of galaxies in front of the Perseus--Pisces
  supercluster, and galaxy chains leading from the Local supercluster
  towards the Coma supercluster \citep{Joeveer:1978a}.  }
\label{fig:wedges}
\end{figure*}

The only known non-baryonic particle was neutrino, thus it is natural
that first neutrinos were considered as dark matter particle
candidates.  The power spectrum of neutrino-dominated dark matter is
cut at small scales due to rapid motion of particles.
Figure~\ref{fig:model} shows results of numerical simulation, based on
such model.  In Fig.~\ref{fig:wedges} the actual distribution of
galaxies, groups and clusters of galaxies is shown.  The comparison
shows that the neutrino-dominated model has no fine structure in the
distribution of galaxies and systems of galaxies.

To avoid these difficulties dissipationless particles heavier than
neutrinos were suggested by \citet{Blumenthal:1982},
\citet{Bond:1982}, and \citet{Peebles:1982}.  Here 
particles like axions, gravitinos or
photinos play the role of dark matter.  These particles were called
cold since free streaming of particles is unimportant and particles
behave as a cold non-dissipative gas.  Numerical models
based on cold dark matter (CDM) represent the fine structure of the
Universe well \citep{Melott:1983}.  The properties of the Cold Dark
Matter model were analysed in detail by \citet{Blumenthal:1984}.

Searches for elementary particles which could serve as candidates for
dark matter particles have been carried out in particle acceleration
centres, so far with no definite results. Thus indirect evidence for
the presence of dark matter has been explored.
One of the recently analysed datasets of interest to investigate
possible effects of dark matter comes from the Fermi Gamma-ray Space
Telescope, launched on June 11, 2008. Its Large Area Telescope (LAT)
can detect gamma rays in an energy interval from about 20~MeV to
300~GeV.  Recently \citet{Weniger:2012} claimed that there is strong
 evidence of a monochromatic gamma-ray line from the Galaxy
centre with an energy $E=130$~GeV  present in the Fermi Large
Area Telescope data.  Soon it was detected that actually there is a
double line spectrum with peaks at energies at 111 and 129 GeV in the
Galactic center.  The double peak-like excess can be interpreted as a
signal of dark matter direct two-body annihilations into two channels
with monochromatic final-state photons.

\begin{figure*}[ht]
    \centering
    \resizebox{.60\columnwidth}{!}{\includegraphics{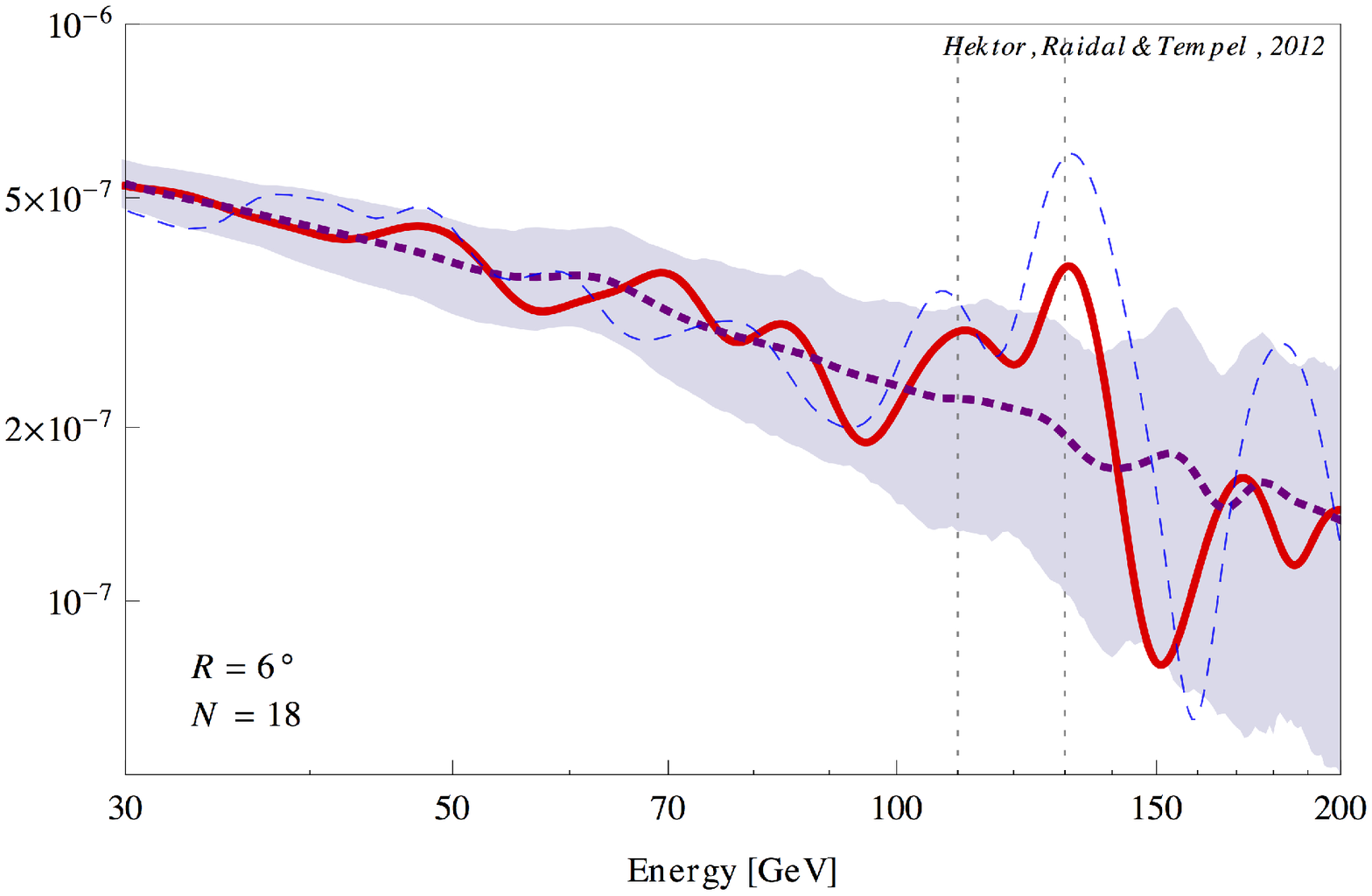}}\\
    \caption{ Gamma-ray spectra for $6^\circ$ regions around the 18
      galaxy clusters as functions of photon energy (solid red 
      curve). The dark dashed line shows a fit to the background
      together with its 95\% error band. The light dashed curve shows
      the reduced signal from the Galactic center for comparison
      \citep{Hektor:2013}.}
    \label{fig:fermi}
\end{figure*}

\citet{Tempel:2012fu,Tempel:2012fk} and \citet{Hektor:2013} analysed
the gamma-ray spectrum of the Galaxy center, as well as of nearby
clusters of galaxies. These authors found from stacked gamma-ray spectra of
clusters a similar double-peak signal.  Figure~\ref{fig:fermi} shows
the observed 110 and 130~GeV excess in Fermi LAT data.  The signal
from galaxy clusters is boosted due to galaxy cluster subhalos. Since
the signal from Galaxy centre and from nearby galaxy clusters shows 
exactly the same double peak structure, the signal must come from the
same physics. Authors conclude: {\em ``The presence of double peak is
  a generic prediction of Dark Matter annihilation pattern in gauge
  theories, corresponding to $\gamma\gamma$ and $\gamma Z$ final
  states. Thus the two seemingly unrelated gamma-ray spectra, from the
  Galactic centre and from the galaxy clusters, favour the particle
  physics origin of the excess over any astrophysics origin''}.  If
these claims are true, this could be a strong evidence that DM
is of particle physics origin, representing a breakthrough both in
cosmology and in particle physics.

The presence of large amounts of matter of unknown origin has given
rise to speculations on the validity of the Newton's law of gravity at
large distances.  One such attempt is Modified Newtonian Dynamics
(MOND), suggested by \citet{Milgrom:1987}.  MOND and other similar
models are able to explain a number of observational data without
assuming the presence of dark matter.

However, there exist several arguments which make these models
unrealistic.  The strongest argument in favour of the presence of
non-baryonic dark matter comes from the CMB data. In the absence of
large amounts of non-baryonic matter during the radiation dominated
era of the evolution of the Universe it would be impossible to get for
the relative amplitude of density fluctuations a value of the order of
$10^{-3}$, needed to form all observed structures.

The other strong argument in favour of the presence of some matter in
addition to ordinary baryonic matter comes also from CMB data.  The
wavenumber of the first acoustic peak in the CMB spectrum is a very
accurate indicator of the total matter/energy density of the
Universe.  Experiments show that with great accuracy the total density
is equal to the critical cosmological density.  On the other hand,
both direct determinations as well as the nucleosynthesis constraints
show that the density of baryonic matter is only about 4~\% of the
critical density.  In other words, there must exist some other forms
of matter/energy than ordinary matter. The other forms are dark matter
and dark energy.  Dark energy causes the acceleration of the Universe,
it was detected by comparison of nearby and distant supernovae by
\citet{Riess:1998} and \citet{Perlmutter:1999}. 

There exist direct observations of the distribution of mass, visible
galaxies and the hot X-ray gas, which cannot be explained in the MOND
framework.  One of such examples is the ``bullet'' cluster 1E 0657-558
\citep{Clowe:2006}.  This is a pair of galaxy clusters, where the
smaller cluster (bullet) has passed the primary cluster almost
tangentially to the line of sight.  Weak gravitational lensing
observations show that the distribution of matter is identical with
the distribution of galaxies.  The hot X-ray gas has been separated by
ram pressure-stripping during the passage.  This separation is only
possible if the mass is in the collisionless component, i.e. in the
non-baryonic dark matter halo, not in the baryonic X-ray gas.

With the discovery of dark energy the buildup of the modern
cosmological paradigm has reached a mature stage. However, the story
of dark matter is not over yet --- we still do not know of what
non-baryonic particles the dark matter is made of, and the nature of
dark energy is completely unknown.

\section{Conclusions}

The main conclusions of the study of dark matter can be formulated as
follows.

\begin{enumerate}
\item{}The discovery of dark matter was the result of combined study
  of galaxies, their populations, and systems of galaxies.
\item{}Dark matter story is a typical paradigm shift. 
Evidence for dark matter has been collected independently in many centres.
\item{}There are two dark matter problems --- dark matter in the Galactic disk,
  and dark matter around galaxies and clusters.
\item{} Dark matter in the Galactic disk is baryonic (faint stars or jupiters). The amount is small.
\item{} Dark matter around galaxies is non-baryonic Cold Dark
  Matter. It constitutes about 0.25 of the critical cosmological density.
\item{}Dark matter is needed to start early enough gravitational
  clustering to form structure. This solves the Big-Bang
  Nucleosynthesis controversy.
\item{}Essential information on the nature of dark matter comes from
  the structure of the cosmic web. The nature of DM particles is still
  unknown.
\end{enumerate}


\end{document}